\documentclass[twocolumn,aps,prl,showpacs]{revtex4}
\usepackage{epsfig}

\begin{document}


\title{Impurity-induced bound states as a signature of pairing symmetry in multiband superconducting CeCu$_{2}$Si$_{2}$}
\author{Dongdong Wang}
\affiliation{Department of Physics, Beijing Jiaotong University, Beijing 100044, China}

\author{Bin Liu}
\email[]{liubin@bjtu.edu.cn}
\affiliation{Department of Physics, Beijing Jiaotong University, Beijing 100044, China}

\author{Min Liu}
\affiliation{Beijing National Laboratory for Condensed Matter Physics and
Institute of Physics, Chinese Academy of Sciences, Beijing 100190, China}

\author{Yi-feng Yang}
\email[]{yifeng@iphy.ac.cn}
\affiliation{Beijing National Laboratory for Condensed Matter Physics and
Institute of Physics, Chinese Academy of Sciences, Beijing 100190, China}
\affiliation{University of Chinese Academy of Sciences, Beijing 100049, China}
\affiliation{Collaborative Innovation Center of Quantum Matter, Beijing 100190, China}

\author{Shiping Feng}
\affiliation{Department of Physics, Beijing Normal University, Beijing 100875, China}

\begin{abstract}

Multiband superconductivity with dominant two-gap features are recently proposed to challenge the earlier accepted nodal $d$-wave pairing in the first unconventional superconductor CeCu$_{2}$Si$_{2}$. Here we obtain multiband Fermi-surface topology of CeCu$_{2}$Si$_{2}$ via first-principles calculations, and study the problem within an effective two hybridization band model including detailed band-structure. Within T-matrix approximation, our calculations reveal that different pairing
candidates could yield qualitatively distinct features characterised by impurity-induced bound states. Except for the nodeless $s^{\pm}$-wave, both loop-nodal $s$-wave and $d$-wave pairings can give rise to intra-gap impurity bound states. In particular, the intra-gap states for the $d_{x^2-y^2}$-wave and loop-nodal $s$-wave are distinguishable and locate either near or far away from the Fermi energy, respectively. These features can be readily verified by high-resolution scanning tunneling microscopy/spectroscopy and provide an unambiguous justification for the ongoing debate about the superconducting gap symmetry of CeCu$_{2}$Si$_{2}$ at ambient pressure.

\end{abstract}

\maketitle
Recently, superconducting (SC) gap symmetry of the heavy-fermion superconductor CeCu$_{2}$Si$_{2}$ has attracted considerable attentions. As the first unconventional superconductor discovered in 1979 \cite{Steglich}, CeCu$_{2}$Si$_{2}$ has long been thought to be a single-band nodal $d$-wave superconductor at ambient pressure, which has been naturally associated with an antiferromagnetic quantum critical point beneath the SC dome and thereby provides a model phase diagram for unconventional superconductivity \cite{Yuan,Steglich1}. The $d$-wave pairing has also been indirectly evidenced by the subsequent specific heat measurement \cite{Bredl,Arndt}, the nuclear magnetic/quadrupole resonance (NMR/NQR) \cite{Ueda,Kitaoka,Fujiwara} and the neutron scattering experimental detections \cite{Stockert}. The only remaining discrepancy seems to be the exact $d$-wave gap structure, either $d_{x^{2}-y^{2}}$ or $d_{xy}$ \cite{Vieyra,Eremin}. However, this view is recently challenged by some latest measurements. In contrast to the nodal $d$-wave pairing, recent specific heat measurement down to very low temperature on CeCu$_{2}$Si$_{2}$ at ambient pressure provides clear evidence of multiband superconductivity with a two-gap structure \cite{Kittaka,Kittaka0,Yamashita}, which has been further convinced by subsequent scanning tunneling microscopy (STM/STS) \cite{Enayat} and London penetration depth measurements \cite{Pang,Takenaka}. Theoretical  calculations using a multiband model afterwards suggest that the superconductivity might be either nodeless $s^{\pm}$-wave \cite{Li}, or loop-nodal $s$-wave \cite{Ikeda}, in contrast to the previous conclusion of $d$-wave based on single band calculations.

Despite that tremendous efforts have been made in the past decades, direct experimental detections of the paring symmetry are still lacking. While the well-known angle-resolved photoemission spectroscopy (ARPES) could provide an exact mapping of the band structures, it is, however, limited by the energy resolution \cite{Fischer}. This makes it difficult to apply for the heavy fermion superconductor, which typically has a SC transition temperature of a few Kelvin and is indiscernible in current ARPES measurement \cite{Knebel,Stewart}. The exact form of gap symmetry in CeCu$_{2}$Si$_{2}$ remains to date elusive. Because of the particular Fermi surface topology of CeCu$_{2}$Si$_{2}$, conventional phase-sensitive measurements cannot be readily applied to detect the pairing symmetry and differentiate the various proposals.

\begin{figure}[tbp]
\begin{center}
\includegraphics[width=1.0\linewidth]{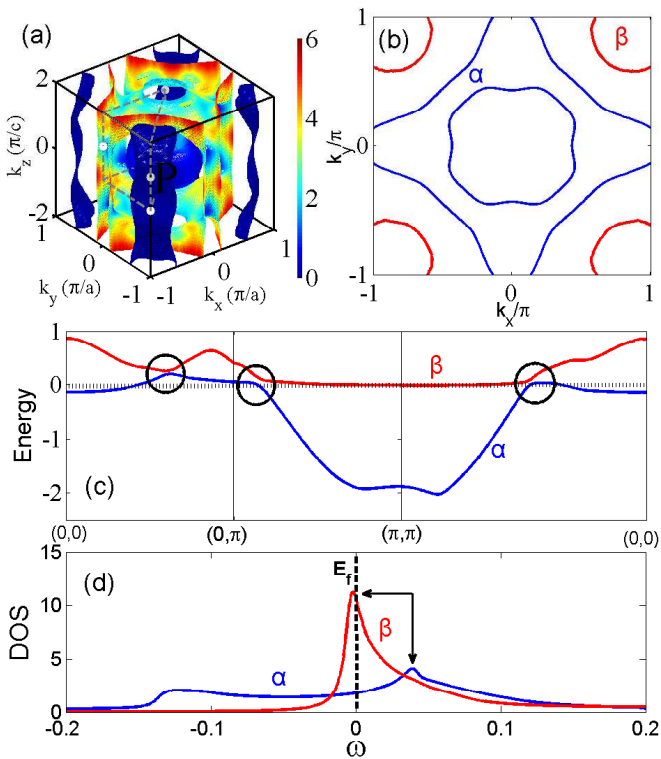}
\end{center}
\caption{(Color online) (a) Calculated 3D Fermi surfaces colored by the Fermi velocity for CeCu$_{2}$Si$_{2}$ at ambient pressure. (b) The 2D FSs of our hybridization-band model. $\beta$ and $\alpha$ denote the corrugated-cylindrical heavy electron sheet and complex hole sheets, respectively. (c) Electronic band structures resulted from hybridization as indicated by the cycles. (d) Density of state of the $\beta$- and $\alpha$-bands in the normal state. The two peaks indicated by arrows are direct consequences of the hybridization. }
\end{figure}

Fortunately, new spectroscopic technique has recently been developed based on high-resolution STM \cite{Aynajian}, where heavy-fermion quasiparticle interference (QPI) could be used to determine the quasiparticle band dispersions at much lower temperature, and measure within $\mu$eV energy scale the detailed {\bf k}-space gap functions. As a result, both QPI and local pair-breaking experiments have been successfully applied to the heavy-fermion superconductor CeCoIn$_5$ to identify its $d_{x^{2}-y^{2}}$ pairing symmetry \cite{Allan,Zhou}. In this paper, we theoretically extend such idea to CeCu$_{2}$Si$_{2}$ as an alternative way to distinguish its SC gap structure. Considering the possibly multiband nature of superconductivity in CeCu$_{2}$Si$_{2}$, we study the problem within an effective model with two hybridization bands obtained from density function calculations with local density approximation taking into account the Coulomb interaction of the Ce $f$-electrons and the spin-orbit coupling (LDA+U). Our results based on T-matrix approximation show that both the loop-nodal $s$-wave and $d$-wave pairing can give rise to impurity-induced intra-gap bound states. For repulsive impurity scattering, the intra-gap states for loop-nodal $s$-wave pairing have energies far away from the Fermi energy irrespective of the impurity scattering strength. This is markedly different from the features shown in the $d$-wave one, where a nearly zero-energy impurity-induced bound state appears for weak impurity scattering. We also reveal that in the case of nodeless $s^{\pm}$-wave, NO intra-gap impurity states, as expected, can be induced by a nonmagnetic impurity. These qualitatively distinct features can be captured in STM and therefore provide a useful guide for  unambiguous experimental justification of these different scenarios to solve this highly debated issue of pairing symmetry in heavy-fermion superconductor CeCu$_{2}$Si$_{2}$.

Since the realistic Fermi surface (FS) topology is crucial for understanding the unconventional superconductivity, we firstly perform band structure calculations for CeCu$_{2}$Si$_{2}$ based on LDA+U. The resulting three-dimensional (3D) FSs for the effective Coulomb interaction $U_{eff}=5\,$eV as shown in Fig. 1(a) includes a corrugated-cylindrical heavy electron sheet around X point, and complex inner hole sheets, in reasonable agreement with previous calculations \cite{Ikeda,Zwicknagl}. The typical 2D FSs of our hybridization band model for $k_{z}$=1.8$\pi$/c is depicted in Fig. 1(b), where the hybridization heavy electron FS around ($\pi$,$\pi$) is denoted as $\beta$-band and the light hole FSs around (0,0) denoted as $\alpha$-band. Fig. 1(c) plots the corresponding band structure, where two heavy-fermion bands clearly appear due to the hybridization from the spin screening of the Kondo lattice at low temperatures as indicated by the cycles, and the extreme flatness of the heavy $\beta$-band near the chemical potential represents typical $f$-characters. As a consequence, the two peaks in the density of state (DOS)  indicated by the arrows in Fig. 1(d) stem from above hybridization between the heavy and light bands. We also note that the heavy $\beta$-band in Fig. 1(d) has a much larger density of state near the Fermi energy, and maybe dominate the low energy properties, which is why  most earlier work were based only on the single-band scenario for studying the unconventional superconductivity in CeCu$_{2}$Si$_{2}$.

We now start with the two hybridization-band model, and introduce a four-component Nambu spinor operator to formulate the bare Green's function in the SC state as
\begin{eqnarray}
\hat{G}^{-1}_{0}({\bf k};i\omega_{n})=i\omega_{n}\hat{1}-\left (\matrix{E^{\alpha}_{\bf k} &\Delta^{\alpha}_{\bf k}
&0 &0\cr \Delta^{\alpha}_{\bf k} &-E^{\alpha}_{\bf k}
&0 &0\cr 0 &0 &E^{\beta}_{\bf k} &\Delta^{\beta}_{\bf k}\cr 0 &0 &\Delta^{\beta}_{\bf k} &-E^{\beta}_{\bf k}\cr}\right),
\end{eqnarray}
where $\omega_n$ is the fermionic Matsubara frequency and $E^{\alpha/\beta}_{\bf k}$ is the dispersion of the $\alpha$- and $\beta$-bands. Since the two hybridization-bands don't overlap, we in this study only discuss the intraband pairing. The considered pairing symmetry includes nodal $d_{x^{2}-y^{2}}$-wave, loop-nodal $s$-wave with $\Delta^{\alpha/\beta}_{\bf k}=\Delta^{\alpha/\beta}\cos k_{x}\cos k_{y}$ and nodeless $s^{\pm}$-wave. Recent specific heat measurement reveals an exponential $T$-dependence which could be fitted by two gaps with a ratio of about 2.5 \cite{Kittaka}, and the magnitude of the gap parameters is not clear and need to be determined from further experiments. We thus choose the same gap symmetry for the two bands with $\Delta^{\beta}/\Delta^{\alpha}\approx 2$ in following numerical calculations, in that this value cannot qualitatively change our conclusions.

The effect of the impurity scattering can be treated within the $T$-matrix approach \cite{Zhu}. For simplicity, we consider only a single nonmagnetic impurity and use the intra-band scattering matrix $\hat{U}$ in the Nambu representation as $U_{ij}=U\delta_{ij}$ for $i=1,3$ and $U_{ij}=-U\delta_{ij}$ for $i=2,4$. Here $U$ is the strength of scattering potential. This yields the full interacting Green's function in real space,
\begin{eqnarray}
\hat{G}(i,j;i\omega_{n})&=&\hat{G}_{0}(i,j;i\omega_{n})\nonumber\\&+&\hat{G}_{0}(i,0;i\omega_{n})\hat{T}(i\omega_{n})\hat{G}_{0}(0,j;i\omega_{n}),
\end{eqnarray}
where $\hat{G}_{0}(i,j;i\omega_{n})=\frac{1}{N}\sum_{\bf k}e^{i\bf k\cdot\bf
{(i-j)}}\hat{G}_{0}({\bf k};i\omega_{n})$ is the the real-space bare Green's function and $\hat{T}(i\omega_{n})=\hat{U}[\hat{1}-\hat{G}_{0}(0,0;i\omega_{n})\hat{U}]^{-1}$ is the the $T$-matrix that has incorporated all the scattering processes. The local density of state (LDOS) can then be evaluated as
\begin{eqnarray}
\rho(i,\omega)=-\frac{1}{\pi}{\rm Im} {\rm Tr} [\hat{G}(i,i;\omega+i0^{\dagger})].
\end{eqnarray}
This quantity is proportional to the local differential tunneling conductance as measured in STM experiments.

\begin{figure}[tbp]
\begin{center}
\includegraphics[width=1.0\linewidth]{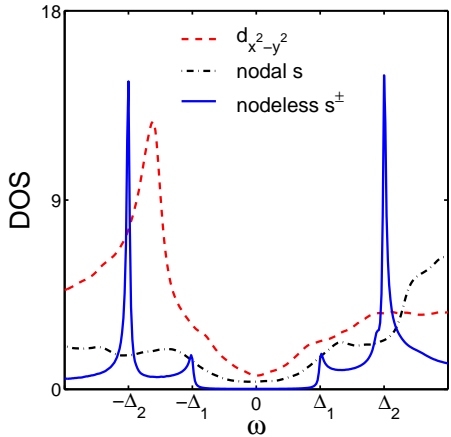}
\end{center}
\caption{(Color online) The density of states in the clean SC state for three gap symmetries, including a nodeless $s^{\pm}$-wave (solid line), a nodal $d_{x^{2}-y^{2}}$-wave (dashed line), and a loop-nodal $s$-wave (dash-dotted line).}
\end{figure}

We first discuss the DOS without impurity in the SC state. As shown in Fig. 2, it exhibits two kinds of coherence peaks corresponding to the maximum SC gaps on the two FSs. The coherence peaks near $\pm \Delta_{1}$ arise from the $\alpha$-band, while those at higher energies $\pm \Delta_{2}$ mainly benefit from the $\beta$-band. For a nodeless $s^{\pm}$-wave, there is no sign change in the SC order parameter within each FS pocket, resulting in the typical U-shape feature in the DOS similar to that in conventional s-wave superconductors \cite{Zhu,Yu}. In contrast, the gap structure of the nodal $d_{x^{2}-y^{2}}$-wave has a sign change and its nodal lines cross both FSs enclosed by the $\alpha-$ and $\beta$-bands, respectively. As a result, the DOS indicated by the dashed line in Fig. 2 shows a V-shape character, in agreement with recent STM measurement \cite{Enayat}. For the SC gap of loop-nodal s-wave with $\Delta^{\alpha/\beta}_{\bf k}=\Delta^{\alpha/\beta}\cos k_{x}\cos k_{y}$, whose gap structure is given in the inset of Fig. 3(a), the nodal lines as indicated by the black dashed line only cut the inner $\alpha$-FS sheet because the radius of the electron $\beta$-FS pocket around ($\pi$,$\pi$) is smaller than $\pi$/2. In this case, the corrugated heavy-electron sheet is fully gapped and only the light-hole sheet has loop nodal points, as proposed in the literature \cite{Ikeda}. The resulting DOS is calculated as a dash-dotted line in Fig. 2.

\begin{figure}[tbp]
\begin{center}
\includegraphics[width=1.05\linewidth]{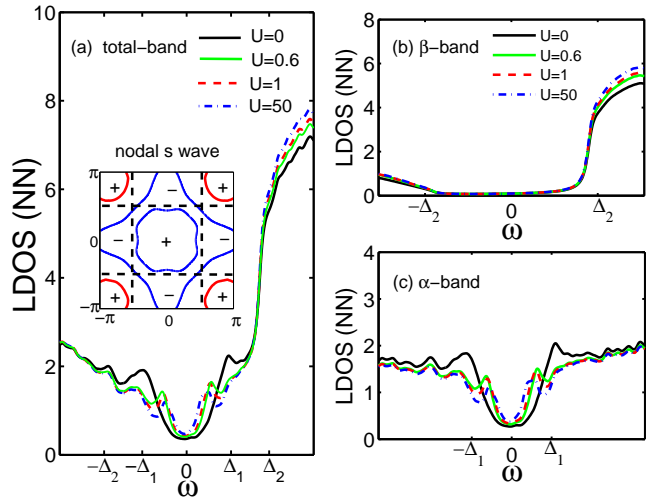}
\end{center}
\caption{(Color online) The LDOS of quasiparticles at the nearest-neighbor (NN) site of the impurity with varying scattering potential for the loop-nodal s-wave with (a) total bands, (b) $\beta$-band, and (c) $\alpha$-band. The inset of (a) is a visualization of the overlap between FSs (solid lines) and the nodal lines (dashed lines) of the gap function; the symbols '+' and '-' denote  positive and negative gap values, respectively.}
\end{figure}

The impurity scattering introduces different local perturbation to the density of states for different SC gap symmetry. Figures 3 and 4 show the LDOS of quasiparticles at the nearest-neighbor (NN) site of the impurity located as the center of the square lattice. To clearly see the intra-gap bound states induced by impurity scattering, we calculate the LDOS from the weak (Born approximation) to strong  (unitary limit) scattering potential strength. In Fig. 3(a), we plot the LDOS by considering the repulsive impurity scattering for loop-nodal s-wave pairing state. The corresponding gap structure is shown in the inset, where the symbols '+' and '-' describe the positive and negative gap values, respectively.  For comparison, the black solid line shows the LDOS in the clean system ($U=0$). In the presence of a single nonmagnetic impurity, one immediately sees that two intra-gap bound states appear, and nearly don't alter the positions with decreasing scattering strength ($U=0.6$ with a unit of meV). Here it is surprising to note that the intra-gap bound peaks are far away from the fermi energy even when the impurity scattering is in the unitary limit ($U= 50$, the dash-dotted line in Fig. 3(a)), in contrast to the situations in high temperature SC cuprates, where the intra-gap bound state is located close to the Fermi energy for strong impurity scattering \cite{Zhu}. Another novel feature is the relatively low spectrum intensities of intra-gap bound states, which can be understood from the nodal gap structure. For loop-nodal s-wave pairing state, the corrugated heavy-electron $\beta$-FS is fully gapped. Therefore, no intra-gap bound states can be induced by local nonmagnetic impurity just as plotted in Fig. 3(b) for $\beta$-band. The obtained intra-gap bound states are completely generated by the inner light hole $\alpha$-FS with loop nodal points. As a result, their peak intensities as seen in Fig. 3(c) are very low due to the smaller DOS of the $\alpha$-band near the Fermi energy as given in Fig. 1(d).

For comparison, the LDOS for the competitive nodeless $s^{\pm}$-wave pairing was calculated and plotted in Fig. 4(a). It is clear that the impurity-induced bound states emerge exactly at the gap edges, which is known as the Yu-Shiba-Rusinov state \cite{Yu}. As expected, no intra-gap bound states can be induced by a nonmagnetic impurity obeying the well-known Anderson's theorem.

Similar calculations have also been performed for the nodal $d_{x^2-y^2}$-wave and the resulting LDOS for varying repulsive impurity scattering are presented in Fig. 4(b). In comparison with the LDOS of the clean system ($U=0$) marked as the black solid line, prominent intra-gap bound states due to impurity pair-breaking occur close to the Fermi energy even for a small value of $U = 0.25$ (dash-dotted line). With increasing $U$, the two peaks move towards the Fermi energy and accidentally merge into a single sharp resonance near zero energy as shown by the red solid line for $U = 0.4$. This nearly zero-energy bound state has already been observed in the $d_{x^2-y^2}$-wave cuprate superconductors \cite{Pan}. In the unitary limit ($U= 50$), two intra-gap bound states reappear and locate symmetrically at the positive and negative energies, but the intensity of the two peaks are different due to the particle-hole asymmetry in the present $\beta$-band.  As shown in the inset of Fig. 4(b), since the nodal lines of the $d_{x^2-y^2}$-wave cross undoubtedly the $\alpha$- and $\beta$-FSs, both FSs contribute to the intra-gap resonance states. However, because of the larger DOS of the corner corrugated heavy-electron $\beta$-band near the Fermi energy, its contribution to the intra-gap states is dominant. Hence the location of the intra-gap states are closer to the Fermi energy, and their intensities are much larger, in contrast to the case of loop-nodal s-wave pairing state considered above. Similar features of the LDOS have also been obtained for attractive impurity scattering, but with a little lower intensity of the intra-gap states (not shown here).

\begin{figure}[tbp]
\begin{center}
\includegraphics[width=1.0\linewidth]{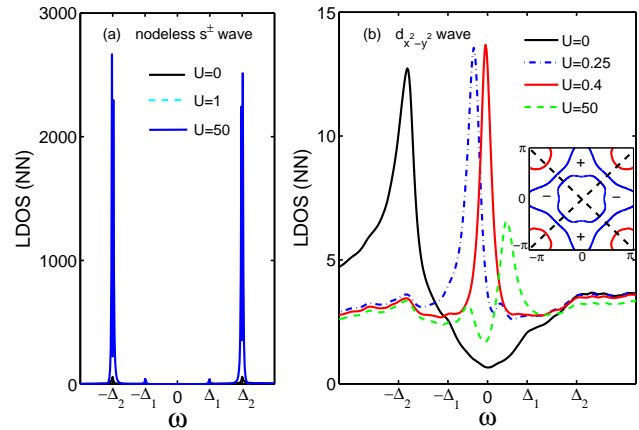}
\end{center}
\caption{(Color online) The LDOS of quasiparticles at NN site for (a) the nodeless $s^{\pm}$-wave and (b) the nodal $d_{x^2-y^2}$-wave pairing symmetries with varying repulsive potential strength. The inset of (b) shows the overlap between the FSs (solid lines) and the nodal lines (dashed lines) of the $d_{x^2-y^2}$ gap function and the symbols '+' and '-' denote positive and negative gap values, respectively.}
\end{figure}

The impurity-induced low-energy quasiparticles will lead to characteristic signatures in scanning tunneling microscopy/spectroscopy, which has recently found quite successful in probing exotic properties in heavy fermion systems such as CeCoIn$_{5}$ \cite{Allan,Zhou}. It is therefore quite appealing to also apply this technique to the heavy fermion superconductor CeCu$_{2}$Si$_{2}$. Given the above qualitative differences, we expect that the tunneling conductance in the STM experiment will provide a promising alternate to distinguish the possible gap structures of the SC gap function for CeCu$_{2}$Si$_{2}$ at ambient pressure.

In conclusion, we study the impurity scattering in the SC state of CeCu$_2$Si$_2$ at ambient pressure using an effective two hybridization-band model derived from the first-principles calculations. The resulting Fermi surfaces include a corrugated-cylindrical heavy electron sheet and a complex hole sheets, thus providing a basis for multiband superconductivity observed in experiment. Our calculations using the T-matrix approach for the effect of impurity scattering on the local electronic structures yield clear differences in the impurity bound states for the $d_{x^2-y^2}$-wave, the nodal $s$-wave and the nodeless $s^\pm$-wave pairings. While there exists no intra-gap bound states for the nodeless $s^\pm$-wave, those for the $d_{x^2-y^2}$-wave and the nodal $s$-wave are distinguishable and locate either near or far away from the Fermi energy, respectively. These qualitatively different features in the LDOS for different pairing states could be measured by future high-resolution STM experiment and thus provide a competitive way to shed light on the SC pairing in CeCu$_2$Si$_2$ at ambient pressure.

This work was supported by the National Natural Science Foundation of China (NSFC) under Grant Nos. 11774025, 11774401 and 11522435. Y.Y. was also supported by the National Key R\&D Program of China (Grant No. 2017YFA0303103) and the Strategic Priority Research Program (B) of the Chinese Academy of Sciences (Grant No. XDB07020200). S.F. was supported by the National Key R\&D Program of China (Grant No. 2016YFA0300304).

\end{document}